\documentstyle[11pt, aaspp4, epsf, rotate]{article}

\def\gamav{$\langle \gamma - 1 \rangle$}

\begin{document}

\title{An analytic function fit to Monte-Carlo X- and $\gamma$-ray spectra
from Thomson thick thermal/nonthermal hybrid plasmas}

\author{M. B\"ottcher\altaffilmark{1,2}, R. Saxena\altaffilmark{3,4},
A. W. Crider\altaffilmark{1,5}, E. P. Liang\altaffilmark{1},
I. A. Smith\altaffilmark{1}, and M. Kusunose\altaffilmark{6}}

\altaffiltext{1}{Physics and Astronomy Department, Rice University, MS 108,
6100 S. Main St., Houston, TX 77005-1892, USA}
\altaffiltext{2}{Chandra Fellow}
\altaffiltext{3}{Clements High School, Sugar Land, TX, USA}
\altaffiltext{4}{current address: UC Berkeley, College of Letters
and Science, Berkeley, CA 94720-2920, USA}
\altaffiltext{5}{Current address: American University, Washington, D.C.}
\altaffiltext{6}{Department of Physics, School of Science,
Kwansei Gakuin University, Nishinomiya 662-8501, Japan}


\centerline{\it Accepted for publication in The Astrophysical Journal 
Supplement Series}

\begin{abstract}
We suggest a simple fitting formula to represent Comptonized
X- and $\gamma$-ray spectra from a hot ($kT_e \gtrsim 10$~keV), 
Thomson thick ($\tau_T \gtrsim 5$) hybrid thermal/nonthermal 
plasma in spherical geometry with homogeneous soft-photon 
injection throughout the Comptonizing region. We have used 
this formula to fit a large data base of Monte-Carlo generated 
photon spectra, and provide correlations between the physical 
parameters of the plasma and the fit parameters of our analytic 
fit function. These allow us to construct Thomson thick
Comptonization spectra without performing computer intensive
Monte Carlo simulations of the high-$\tau_{\rm T}$ hybrid-plasma
Comptonization problem. Our formulae can easily be used in data 
analysis packages such as XSPEC, thus rendering rapid $\chi^2$ 
fitting of such spectra to real data feasible. 
\end{abstract}

\section{Introduction}
Compton upscattering of soft (optical, UV or soft X-ray) radiation
by hot ($kT_e \gtrsim 10$~keV) plasma is believed to play an important
role in the formation of high-energy (hard X-ray and $\gamma$-ray)
spectra of many astrophysical objects. Some examples are Galactic 
X-ray binaries (for recent reviews see \cite{tl95,liang98}), 
supernova remnants (e.g., \cite{hillas98,aharonian99}), Seyfert 
galaxies (for recent reviews see \cite{mush93,zdz99}), 
blazars (\cite{maraschi92,dermer92}), the intergalactic medium 
(\cite{sz80,reph95}), and possibly also $\gamma$-ray bursts 
(\cite{liang97,liang99}). 

Comptonization in purely thermal plasmas is fairly well 
understood, and analytical solutions for the emerging hard X-ray and 
$\gamma$-ray spectra for a wide range of Thomson depths and plasma 
temperatures have been developed (\cite{st80,titarchuk94,tilu95,ht95}). 
However, the currently known solutions can not be applied to 
hybrid thermal/nonthermal or even purely nonthermal plasmas, 
which are believed to be the primary source of hard X-ray 
radiation in many astrophysical objects. For example, Li, 
Kusunose, \& Liang (\markcite{li96}1996) have shown that
the high-energy spectrum of Cyg~X-1 can be plausibly reproduced 
by Comptonization in a quasi-thermal plasma with a small fraction 
of electrons in a nonthermal tail. Thermal/nonthermal hybrid models
have also been discussed in the context of the high-energy spectra
of Seyfert galaxies, e.g. NGC~4151 (\cite{zdz94,zdz96}). 
The development of such nonthermal tails has been confirmed 
by B\"ottcher \& Liang (\markcite{bl01}2001) by detailed 
Monte-Carlo/Fokker-Planck simulations of radiation transport
and electron dynamics in magnetized hot plasmas, e.g. in 
accretion flows onto compact objects.

We need to point out that the combination of soft- to medium-energy
X-ray spectra (typically integrated over a few ksec) with hard X-ray
/ soft $\gamma$-ray OSSE spectra (typically integrated over hundreds of
ksec or even several days) could introduce artificial spectral signatures 
due to the superposition of thermal Comptonization spectra from different 
emission regions, possibly at different times. However, the detailed 
Fokker-Planck simulations of the dynamics of electron acceleration and 
cooling in coronal plasmas in the vicinity of accreting compact objects 
(e.g., \cite{li96,bl01}) clearly demonstrate that nonthermal tails 
in the electron distributions are very likely to be produced in 
such environments. It is thus desirable to have simple, analytical 
expressions for the Comptonization spectra arising in this case.

For the general case of arbitrary Thomson depth, average electron
energy (i.e. temperature), nonthermal fraction of electrons in the
hybrid plasma, and nonthermal electron spectral parameters, the 
most reliable and often fastest method for computing Comptonization 
spectra is the Monte-Carlo method (\cite{poz83}). However, 
Monte-Carlo simulations of Comptonization problems become 
extremely time consuming in the case of very high Thomson depth, 
$\tau_{\rm T}$, since the number of scatterings that need to be simulated, 
increases $\propto \tau_{\rm T}^2$. To date, there are no simple, analytic 
approximations to the problem of optically thick Comptonization in hot, 
thermal/nonthermal hybrid plasmas. This has made $\chi^2$  minimization 
fitting of such models to real data infeasible until now. 

Using our linear Monte-Carlo Comptonization codes, we have built 
up a data base of over 300 Comptonization spectra from hot, Thomson 
thick, thermal/nonthermal hybrid plasmas with different Thomson 
depths, temperatures, nonthermal electron fractions, spectral indices
and maximum electron energies of nonthermal electrons. In this paper,
we propose a simple analytic representation, consisting of an exponentially
cut-off power-law plus a smoothly connected double-power-law (\cite{band93})
to fit all individual spectra in our data base. We find correlations
between the physical parameters of the Monte-Carlo calculations and the
parameters of our analytical representation, which can be used to construct
Comptonization spectra of high-$\tau_{\rm T}$ plasmas with arbitrary
nonthermal electron fractions, without having to perform time-consuming
Monte-Carlo simulations. Our results are specifically optimized for 
very high Thomson depths, $\tau_T \gtrsim 5$, and are valid for electron
temperatures $kT_e \gtrsim 10$~keV, since we assume the Comptonizing
plasma to be fully ionized. For lower temperatures, effects of photoelectric
absorption, recombination, and other atomic processes, which have been 
neglected in the current work, must be included.

\section{Model setup and analytic representation of the Comptonization spectra}

The physical scenario underlying the Comptonization problem at 
hand, consists of a spherical, homogeneous region of hot ($kT_e 
\gtrsim 10$~keV), thermal/nonthermal hybrid plasma with radial 
Thomson depth $\tau_{\rm T}$. The thermal/nonthermal hybrid
distribution function of electron energies is given by

\begin{equation}
f_e (\gamma) = \cases{ N_{\rm th} \, \gamma^2 \, \beta \; e^{- \gamma / \Theta_e}
& for $1 \le \gamma < \gamma_1$ \cr
N_{\rm nt} \, \gamma^{-p} & for $\gamma_1 \le \gamma \le \gamma_2$ \cr}
\label{el_distribution}
\end{equation}
Here, the normalization factors $N_{\rm th}$ and $N_{\rm nt}$ and the
transition energy $\gamma_1$ are determined by the normalization,
$\int_1^{\infty} f_e (\gamma) \, d\gamma = 1$, by the requirement of
continuity of the distribution function at $\gamma_1$, and by the
fraction $a_{\rm mxwl}$ of thermal particles, i. e. $\int_1^{\gamma_1}
f_e (\gamma) \, d\gamma = a_{\rm mxwl}$. 

Soft photons with characteristic energy $E_s \ll 1$~keV 
are injected uniformly throughout the Comptonizing region 
(we assume that they have a thermal blackbody spectrum with
$kT_r \ll 1$~keV; the specific shape of the soft photon 
distribution is irrelevant for multiple-Compton-scattering
problems). The Comptonizing plasma is assumed to be fully
ionized. The Comptonization spectra are calculated with our 
Monte-Carlo codes, which use the full Klein-Nishina cross
section for Compton scattering. $\gamma\gamma$ absorption, 
pair production, and other induced processes (e.g., 
free-free absorption and induced Compton scattering events) 
have been neglected in our Monte-Carlo simulations.

We have created a data base of over 300 simulations, spanning values
of $\tau_{\rm T} \le 28$, $5 \, {\rm keV} \le kT_e \le 200$~keV, $0.2 
\le a_{\rm mxwl} \le 1$, $2.5 \le p \le 6.5$, and $10 \le \gamma_2 
\le 10^3$.

We here propose a fit function to optically thick Comptonization
spectra from thermal/nonthermal hybrid plasmas as the sum of an
exponentially cut-off power-law plus a Band GRB function (defined
here, for convenience, in energy flux units to match the units of
the output spectra of our Monte-Carlo simulations):

\begin{eqnarray}
F_{\rm E} (E)&=&N_{\rm PL} \, E^{- \Gamma} e^{-E / E_0} \cr\cr
&+&A \cdot 
\cases{ E^{\delta} e^{-E / E_0} & for $E < (\delta - \epsilon) E_0$ \cr\cr
        B \, E^{\epsilon} & for $E > (\delta - \epsilon) E_0$ \cr}
\label{f_definition}
\end{eqnarray}
with $B \equiv \bigl( [\delta - \epsilon] E_0 \bigr)^{\delta - \epsilon}
\, e^{\epsilon - \delta}$ and $E$ in keV. $F_{\rm E}$ is the energy flux 
in units of ergs~cm$^{-2}$~s$^{-1}$~keV$^{-1}$.

\begin{figure}
\centering
\epsfxsize=0.9\hsize
\epsffile{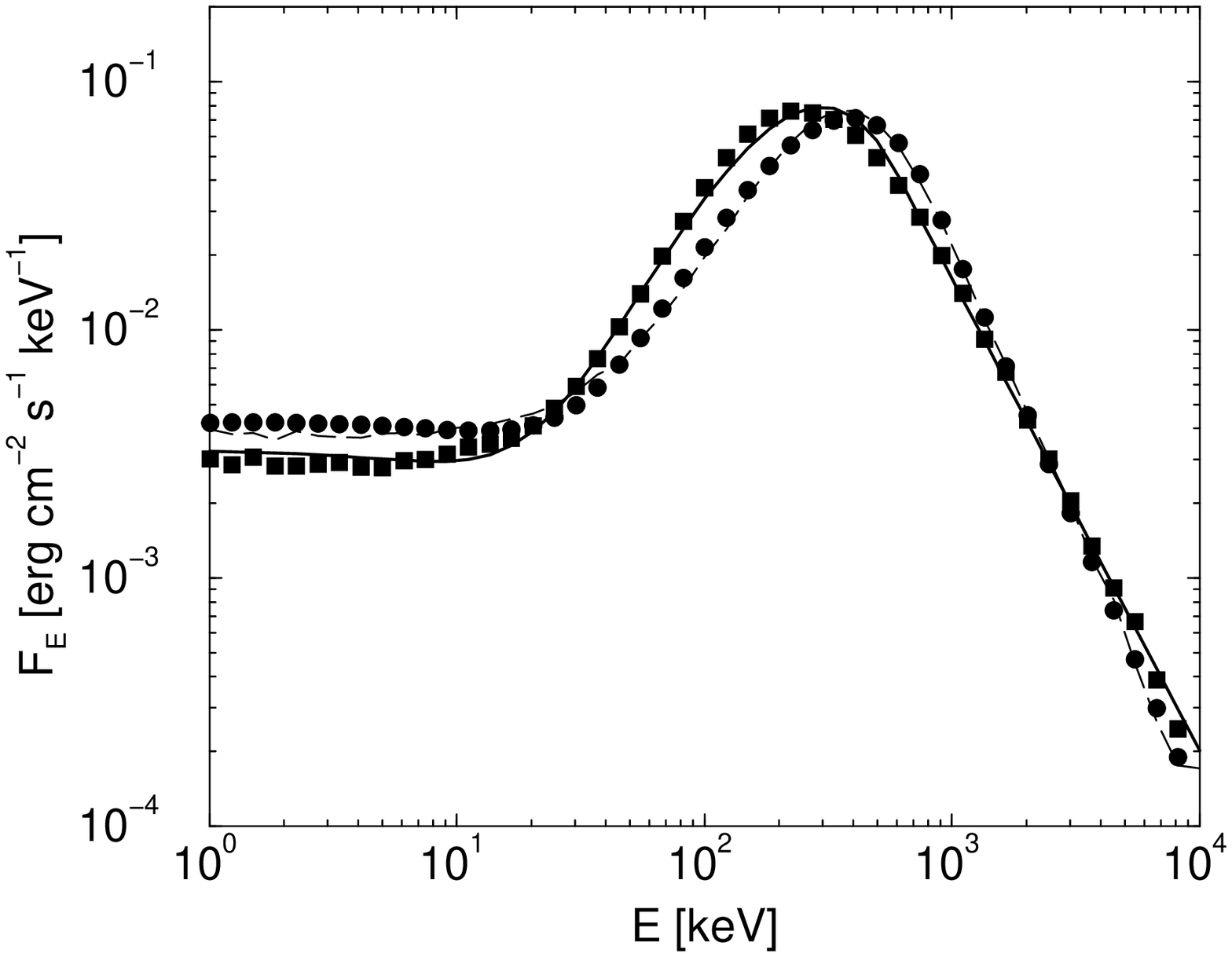}
\caption{Two examples of fits of our model function (solid and dashed
curves) to the Monte-Carlo simulated energy spectrum (filled symbols). 
Parameters for the Monte-Carlo simulation (filled circles) are: 
$\tau_{\rm T} = 16$, $kT_e = 80$~keV, $a_{\rm mxwl} = 0.5$, $p = 3.5$, 
$\gamma_2 = 10^3$, $kT_s = 0.5$~eV. Parameters of the analytic fit 
(dashed curve) are: $\Gamma = -0.035$, $\delta = 1.98$, $\epsilon = -2.27$, 
$E_0 = 196$~keV, $A/N_{pl} = 8.84\times 10^{-4}$. Parameters for the
second MC simulation (filled squares) are: $\tau_{\rm T} = 28$, 
$kT_e = 40$~keV, $a_{\rm mxwl} = 0.7$, $p = 3.5$, $\gamma_2 = 10^3$, 
$kT_s = 0.5$~eV. Parameters of the analytic fit (solid curve): 
$\Gamma = -0.029$, $\delta = 2.01$, $\epsilon = -1.90$, $E_0 = 149$~keV, 
$A/N_{pl} = 1.87\times 10^{-3}$.}
\label{fit_example1}
\end{figure}

Apart from the overall normalization, there are 5 parameters in this 
fit function: the spectral indices $\Gamma$, $\delta$, and $\epsilon$, 
the turnover energy $E_0$ (corresponding to a $\nu F_{\nu}$ peak 
energy of $E_{\rm pk} = [1 + \delta] \, E_0$), and the relative 
normalization of the two components, $A / N_{\rm PL}$. We have developed 
a code using a combination of a coarse grid in the 6-dimensional parameter 
space with a $\chi^2$ forward folding method to fit this function to our 
simulated Compton spectra. Fig. \ref{fit_example1} shows two typical examples
of the resulting fits.

\section{Correlations between physical and fit parameters}

We have fitted the function (\ref{f_definition}) to the energy
range $1 \, {\rm keV} \le E \le 10$~MeV of all simulated Comptonization 
spectra in our data base. In order to be able to use our fit function 
for physically meaningful fitting, one has to establish a unique 
correlation between the physical parameters of the Monte-Carlo simulation 
and the best-fit parameters of our fit function. First of all, we note 
that in the low-energy range, dominated by the power-law part of the
spectrum, $F_{\rm E} \propto E^{- \Gamma}$, the influence of the nonthermal
population is negligible, and we can use the standard result for thermal
plasmas in spherical geometry in the case of homogeneous photon
injection throughout the Comptonizing region:

\begin{equation}
\Gamma = - {3 \over 2} \left( 1 - \sqrt{1 + {4 \, \pi^2 \over 27 \,
\Theta_e \, (\tau_{\rm T} + 2/3)^2}} \right)
\label{Gamma}
\end{equation}
(\cite{st80}) where $\Theta_e = kT_e / (m_e c^2)$. This relation is
valid in the limit of high Thomson depth, $\tau_T \gtrsim 5$, and
moderate electron temperature, $kT_e \lesssim 250$~keV (\cite{st80,poz83}). 
A fully analytical solution for the power-law index $\Gamma$ for all 
values of $\tau_{\rm T}$ and $\Theta_e$ has been derived by Titarchuk 
\& Ljubarskij (\markcite{tilu95}1995). In the parameter range in which
our analytical representation is applicable, the simple expression
(\ref{Gamma}) is sufficiently accurate.

\begin{figure}
\centering
\epsfxsize=0.9\hsize
\epsffile{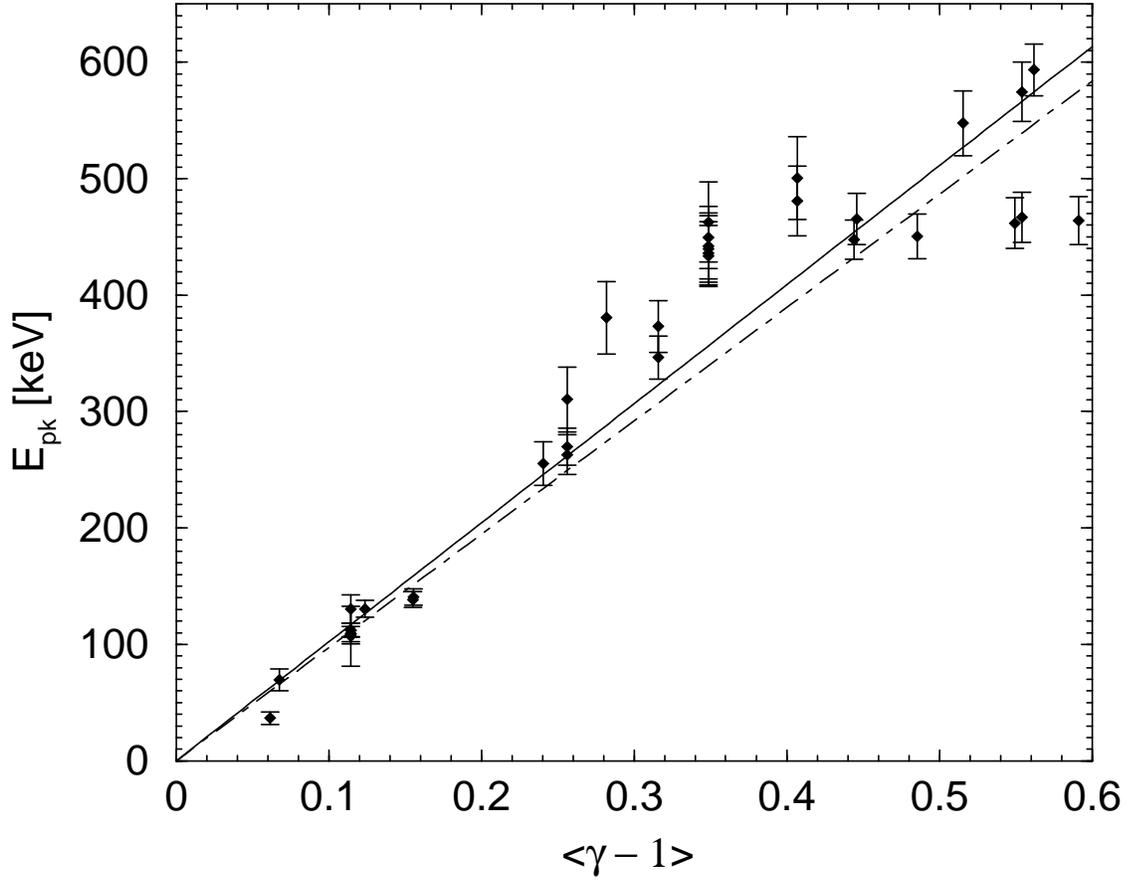}
\caption{Correlation between the average electron energy in the plasma,
\gamav, and the $\nu F_{\nu}$ peak energy, $E_{\rm pk}$, compared to the
expected correlation, $E_{\rm pk} = 1022 \langle \gamma - 1 \rangle$~keV
(solid line). The best-fit linear slope is $973$~keV (dot-dashed line).}
\label{Epk_gminus1}
\end{figure}

A second, obvious correlation exists between the $\nu F_{\nu}$ peak of an 
optically thick Comptonization spectrum and the average particle energy. 
In the thermal case, the Wien spectrum peaks at $E_{\rm pk} \approx 3 \, 
kT_e$. However, near the $\nu F_{\nu}$ peak of the spectrum, the influence 
of the nonthermal particle population can become important. Generalizing 
the $E_{\rm pk}$ vs. $kT_e$ correlation, one would expect that $E_{\rm pk} 
\approx 2 \langle \gamma - 1 \rangle \, m_e c^2$, where the brackets denote 
the ensemble average. We find that this generally provides a reasonable 
fit to the simulated $E_{\rm pk}$ values. However, in some cases
(especially for hard power-law tails, $p \lesssim 3$), the onset of
the nonthermal population (at $\gamma_1$) is at too high an energy
and the normalization of the nonthermal population is too small for 
nonthermal electrons to make a strong contribution to photons scattered
into the Wien peak. We find significant deviations from the above
$E_{\rm pk}$ vs. \gamav\ correlation for values of $\gamma_1 >
6 \, \Theta_e + 1$. An approximate correction for this effect can
be found in the following empirical correlation which provides 
a reasonable description of the simulated $E_{\rm pk}$ values:

\begin{equation}
E_{\rm pk} = \cases{ 1.022 \, \langle \gamma - 1 \rangle \; {\rm MeV}
& for $\gamma_1 < 6 \, \Theta_e + 1$ \cr\cr
{a_{\rm mxwl} \langle\gamma - 1\rangle_{\rm th} + (1 - a_{\rm mxwl})
\langle\gamma - 1\rangle_{\rm nt} \, 2^{-f/8} \over a_{\rm mxwl}
+ (1 - a_{\rm mxwl}) \, 2^{-f/8}} \; {\rm MeV} & for $\gamma_1 \ge 
6 \, \Theta_e + 1$ \cr}
\label{Epk}
\end{equation}
where $f \equiv 2 (\gamma_1 - 1)/\Theta_e$. The terms \gamav$_{\rm th}$
and \gamav$_{\rm nt}$ denote the average kinetic energies (normalized to
$m_e c^2$) of electrons in the thermal and nonthermal portion of the 
electron distribution, respectively. The $E_{\rm pk}$ vs. $\gamma - 1$ 
correlation for those cases with $\gamma_1 < 6 \, \Theta_e + 1$ is 
illustrated in Fig. \ref{Epk_gminus1}. We note that in most physically
relevant cases, we expect the power-law index of nonthermal electrons
to be $p \gtrsim 3$ since we do not expect a positive $\nu F_{\nu}$ 
spectral slope at high energies due to the nonthermal portion of the 
hybrid electron spectra. In the case $p \gtrsim 3$, the upper branch 
of Eq. (\ref{Epk}) always yields satisfactory results.

\begin{figure}
\centering
\epsfxsize=0.9\hsize
\epsffile{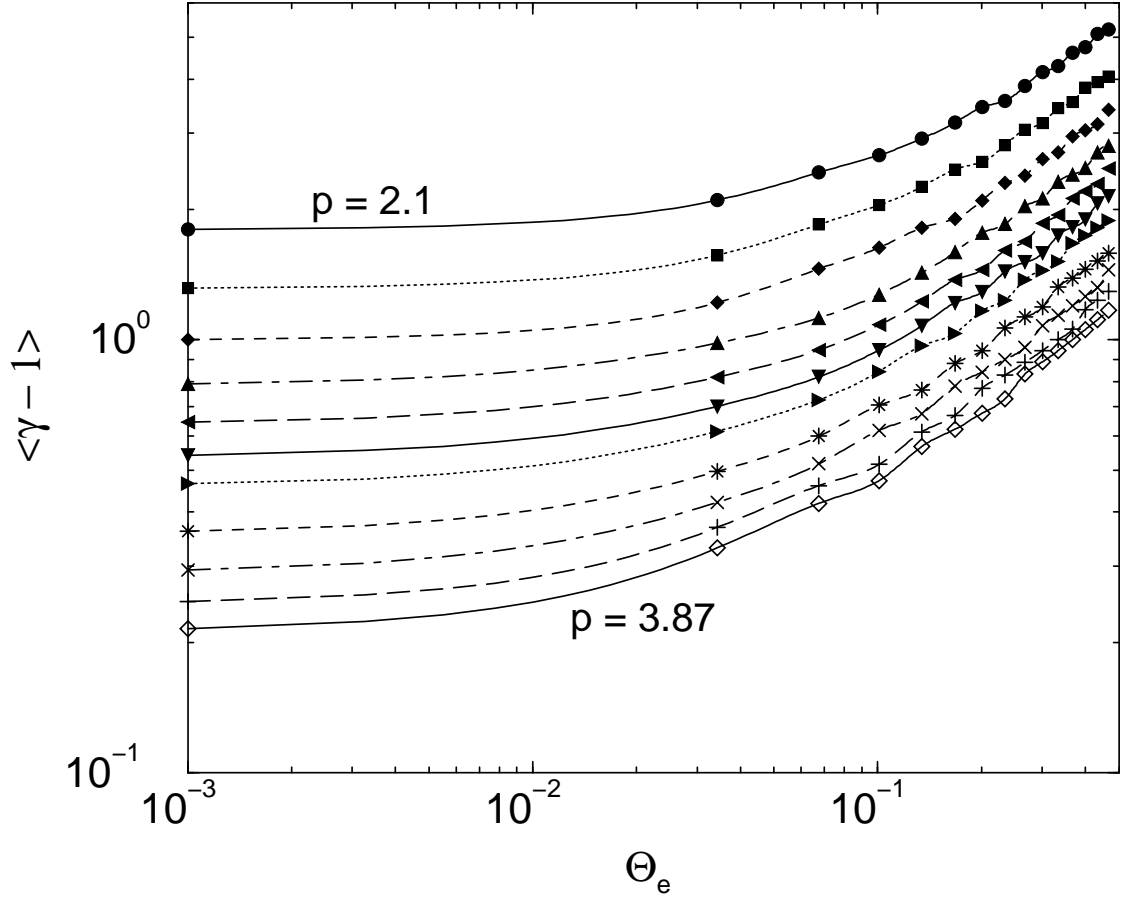}
\caption{Dependence of \gamav\ on plasma temperature and nonthermal
spectral index $p$; example for $a_{\rm mxwl} = 0.6$, $\gamma_2 =
10^3$. From top to bottom, the nonthermal spectral indices are
$p = 2.1$, 2.23, 2.35, 2.48, 2.61, 2.73, 2.86, 3.11, 3.37, 3.62,
and 3.87.}
\label{gminus1_graph}
\end{figure}

The value of \gamav\ depends on the temperature of the thermal population 
and the parameters of the nonthermal population in a non-trivial way. 
We find that \gamav\ can be parametrized rather accurately by the 
fitting function

\begin{equation}
\langle \gamma - 1 \rangle \approx K_1 \, \Theta_e^{\kappa_1} +
K_2 \, e^{\Theta_e / \kappa_2}.
\label{gamav_fit}
\end{equation}
To facilitate the use of our fitting formulae, we list in Table
\ref{gm1_table} the relevant fit parameters $K_1$, $K_2$, $\kappa_1$, 
and $\kappa_2$ for a representative value of $\gamma_2 = 10^3$. 
The dependence of \gamav\ on the thermal plasma temperature and 
the nonthermal power-law index $p$ is illustrated in Fig. 
\ref{gminus1_graph}.

\begin{figure}
\centering
\epsfxsize=0.9\hsize
\epsffile{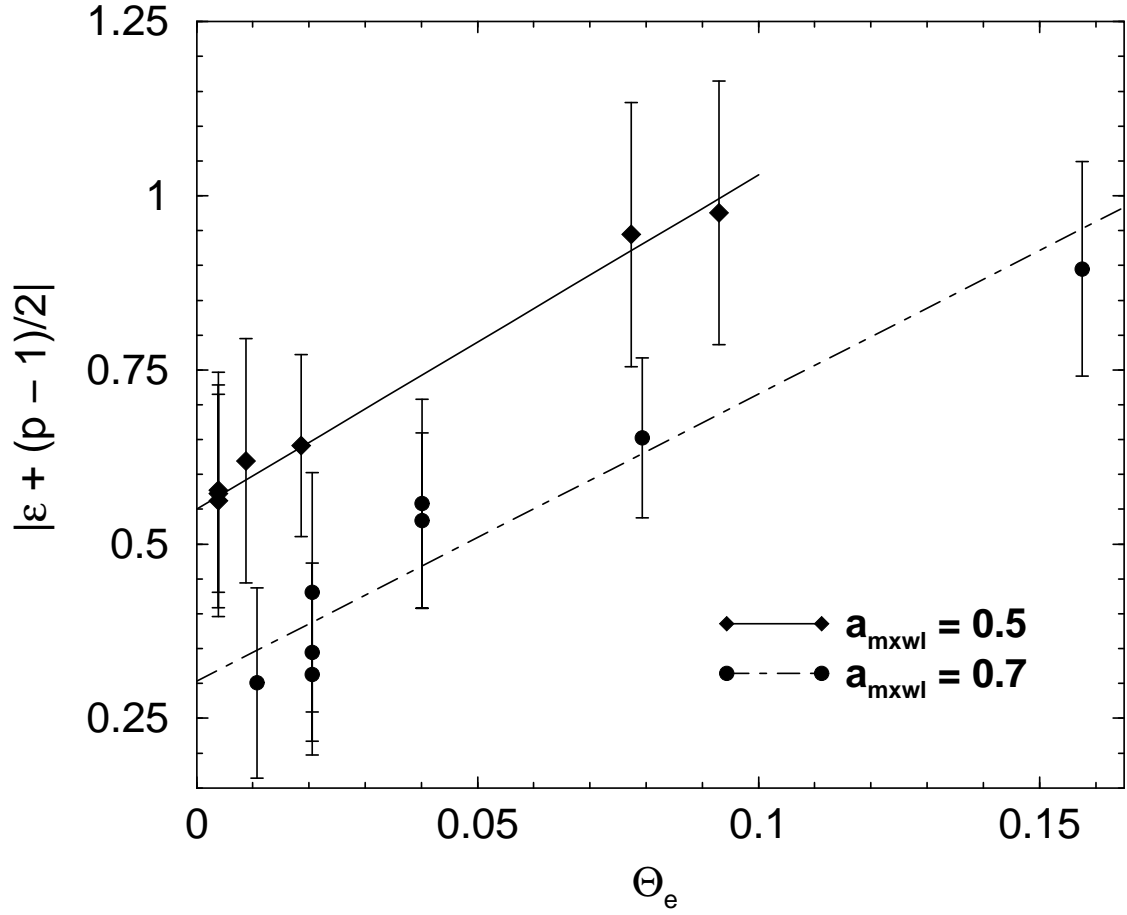}
\caption{Correlation between the correction to the high-energy spectral
index $\epsilon$ and the plasma temperature for two different values of
the Maxwellian plasma fraction $a_{\rm mxwl}$.}
\label{epsilon_graph}
\end{figure}

The connections of the remaining fit parameters, $\epsilon$,
$\delta$, and $A / N_{\rm PL}$ with the physical parameters
of the Comptonizing plasma, are less obvious, and the
correlations which we have found on the basis of the fit 
results to our data base, are purely empirical. In an optically 
thin ($\tau_{\rm T} \ll 1$), purely nonthermal plasma, the 
high-energy spectral index $\epsilon$ would be related to 
the non-thermal particle index $p$ through $\epsilon = - 
(p - 1)/2$ (e.g., \cite{rl79}). However, the presence of 
the thermal population as well as the effect of multiple 
Compton scatterings lead to a systematic distortion of 
this spectral shape. We find a correction to $\epsilon$ due 
to these effects, which can be parametrized as

\begin{equation}
\epsilon = - {p - 1 \over 2} - (a_{\epsilon} + b_{\epsilon} \Theta_e)
\label{beta}
\end{equation}
with
\begin{eqnarray}
a_{\epsilon} &= &0.78 \, e^{-(1.49 \, a_{\rm mxwl})^3} \\
b_{\epsilon} &= &3.25 \, a_{\rm mxwl}^{-0.7}
\label{beta_params}
\end{eqnarray}
Two examples of the correlation (\ref{beta}) are shown in 
Fig. \ref{epsilon_graph}.

The low-energy power-law slope of the Wien hump, added to the 
low-energy power-law (the $F_{\rm E} \propto E^{\delta}$ part
of our fitting function), approaches $\delta \to 2$ for very 
high Thomson depth. We find that the correlation between $\delta$ 
and $\tau_{\rm T}$ can be well parametrized by a functional form

\begin{equation}
\delta = 2 {\tau_{\rm T}^{\eta (\Theta_e)} \over
[\tau_0 (\Theta_e)]^{\eta (\Theta_e)} 
+ \tau_{\rm T}^{\eta (\Theta_e)}}
\label{alpha}
\end{equation}
which is virtually independent of $a_{\rm mxwl}$ and the
specifications of the nonthermal population. The parameters
in Eq. \ref{alpha} depend on the thermal electron temperature 
as

\begin{eqnarray}
\tau_0 (\Theta_e) &= &5.6 \, \Theta_e^{0.13} \\
\eta (\Theta_e) &= &2 \, \left( \ln[200 \, \Theta_e] \right)^{1.7}
\label{alpha_params}
\end{eqnarray}

\begin{figure}
\centering
\epsfxsize=0.9\hsize
\epsffile{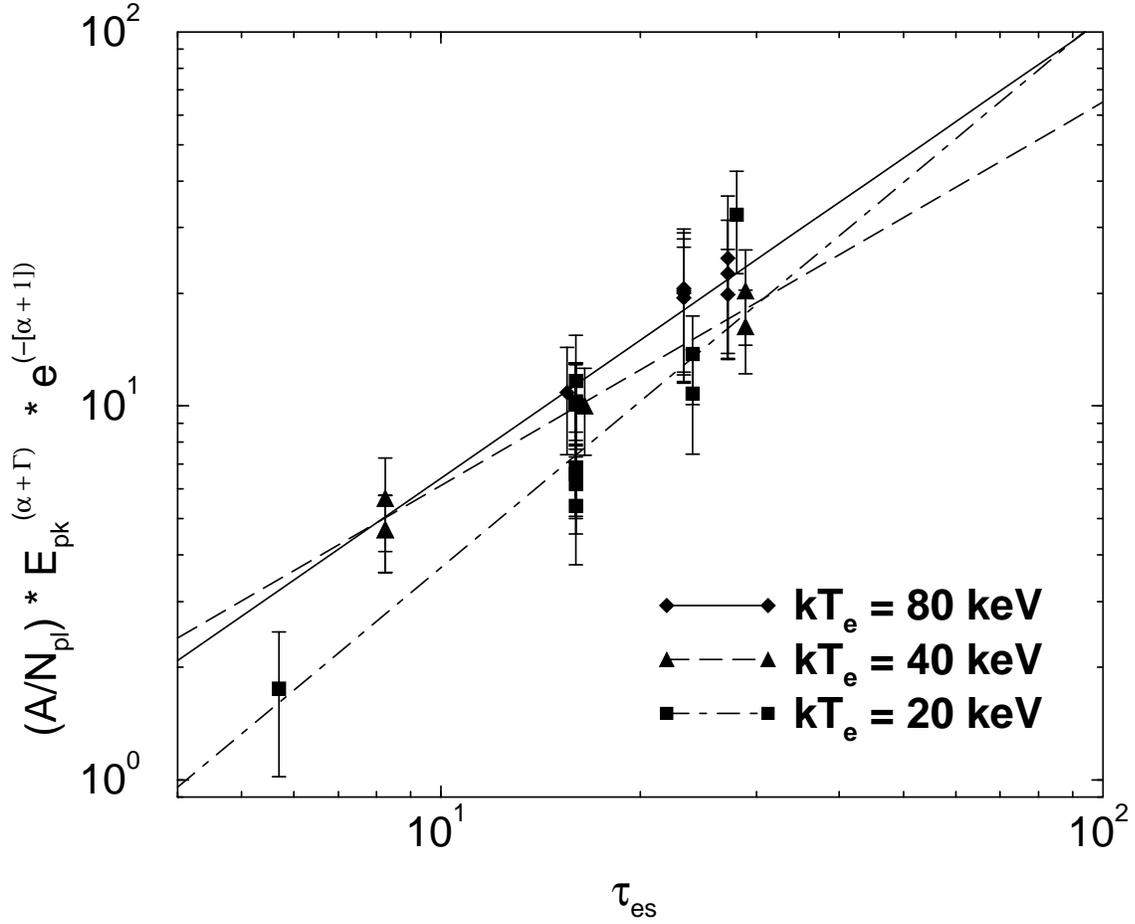}
\caption{Empirical correlation between the relative power in
the power-law to the Wien hump (Band function part of the photon 
spectrum) and the Thomson depth $\tau_{\rm T}$ for various
electron temperatures (with various values of $a_{\rm mxwl}$
each, which does not appear to have a significant influence).
The lines are the best-fit power-laws.}
\label{ANpl_graph}
\end{figure}

Finally, we expect the ratio of power in the Wien hump (Band
function section of our fit function) to the power in the 
low-energy power-law part of the Comptonization spectrum 
to be positively correlated with the plasma temperature $\Theta_e$ 
and Thomson depth $\tau_{\rm T}$. The above mentioned power 
ratio can be expressed as the term 

\begin{equation}
C_{A/N_{\rm PL}} \equiv (A /N_{\rm PL}) \, E_{\rm pk}^{\delta
+ \Gamma} \, e^{- (\delta + 1)}. 
\label{C_A_Npl_def}
\end{equation}
To first order, this power ratio should not strongly depend 
on the specifications of the non-thermal component. Fig. 
\ref{ANpl_graph} shows some examples of the correlation
between $C_{A/N_{\rm PL}}$ and $\tau_{\rm T}$ for various
plasma temperatures, where we have neglected any dependence
on $a_{\rm mxwl}$ and $p$. We find that this correlation 
can be parametrized as

\begin{equation}
{C'}_{A/N_{\rm PL}} = C_0 \, \left( {\tau_{\rm T} \over 20} 
\right)^{a_C} + C_1
\label{C_A_Npl_corr}
\end{equation}
with 

\begin{eqnarray}
C_0 &= &16 \, \Theta_e^{0.1} \\
C_1 &= &5 \\
a_C &= &\Theta_e^{0.05}
\label{C_A_Npl_params}
\end{eqnarray}
However, comparing the analytic spectra calculated using 
Eq. (\ref{C_A_Npl_corr}) with the Monte-Carlo generated spectra, 
we do find significant $a_{\rm mxwl}$-dependent deviations from 
the best-fit values of $A / N_{\rm PL}$. We thus introduce a 
correction ${C''}_{A/N_{\rm PL}}$ such that 

\begin{equation}
C_{A/N_{\rm PL}} = {C'}_{A/N_{\rm PL}} \cdot {C''}_{A/N_{\rm PL}}
\label{ANpl_correction}
\end{equation}
and find this correction term as

\begin{equation}
{C''}_{A/N_{\rm PL}} = (2.1 \cdot 10^{-3} \Theta_e^{-2.3} + 1.8) \,
a_{\rm mxwl}^{-2.1} \, \tau_T^{- \lambda (\Theta_e)}
\label{correctionterm}
\end{equation}
with
\begin{equation}
\lambda (\Theta_e) = 2.8 \cdot 10^{-2} \, \Theta_e^{-0.95}.
\label{lambda}
\end{equation}
Since we know $E_{\rm pk}$, $\delta$ and $\Gamma$ from Eqs.
(\ref{Epk}), (\ref{alpha}) and (\ref{Gamma}), we can easily
invert Eq. (\ref{C_A_Npl_def}) to find the fit parameter
$A / N_{\rm PL}$. 

\section{Summary and Conclusions}

\begin{figure}
\centering
\epsfxsize=0.9\hsize
\epsffile{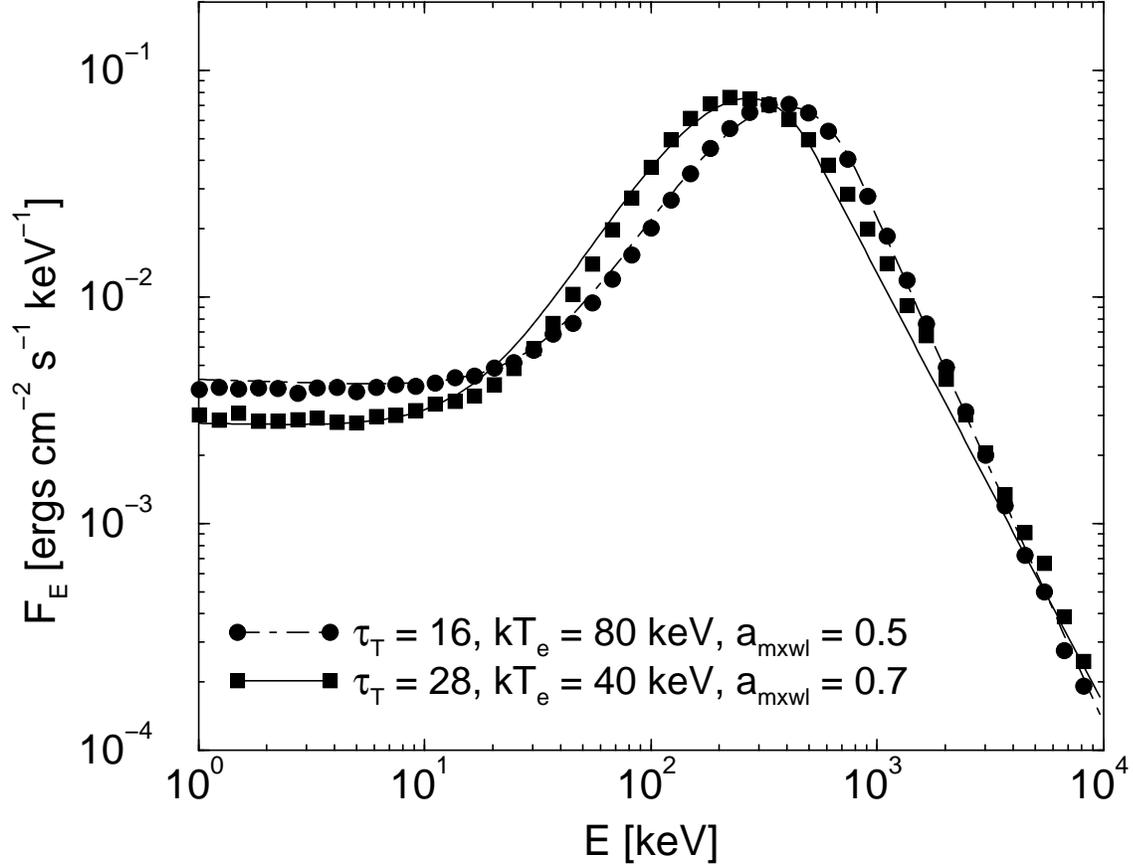}
\caption{Two typical examples of the analytical representation
for Thomson thick Comptonization spectra, using Eqs. (\ref{Gamma}),
(\ref{alpha}), (\ref{beta}), (\ref{Epk}), and (\ref{C_A_Npl_corr})
(solid and dot-dashed curves), compared to the Monte-Carlo simulated 
output spectra (symbols; same simulations as shownin Fig. \ref{fit_example1}). 
In both simulations, $p = 3.5$ and $\gamma_2 = 10^3$ was used.}
\label{fit_example2}
\end{figure}

We have developed an analytical representation of Comptonization 
spectra from optically thick, hot, thermal/nonthermal hybrid 
plasmas in spherical geometry with homogeneous soft-photon
injection. Starting from the physical parameters of the problem
at hand, one can easily use Eqs. (\ref{Gamma}), (\ref{alpha}), 
(\ref{beta}), (\ref{Epk}), and (\ref{C_A_Npl_corr}) to construct
the resulting photon spectrum. Thus, our representation can be
used in $\chi^2$ minimization software packages to scan through
the physical parameter space of the thermal/nonthermal hybrid
plasma and fit the emerging Comptonization spectra to observed
photon spectra. Fig. \ref{fit_example2} shows two representative
examples of the analytical spectra, using the correlations presented
in the previous section, and the output spectra from the corresponding
Monte-Carlo simulations. 

Our results are valid for electron temperatures $10 \, {\rm keV} 
\lesssim kT_e \lesssim 150$~keV, Thomson depths $\tau_{\rm T} \gtrsim 5$, 
Maxwellian electron fractions $0.5 \lesssim a_{\rm mxwl} \lesssim 0.95$
(for purely thermal plasmas, $a_{\rm mxwl} = 1$, the analytical 
solutions of Hua \& Titarchuk [\markcite{ht95}1995] may be used), 
nonthermal electron spectral indices $p \gtrsim 3$, and seed photon 
energies $E_s \ll 1$~keV. The results have been obtained for
homogeneous photon injection in spherical geometry and should not
be used for other geometries.

\section*{Acknowledgments}

The work of M.B. is supported by NASA through Chandra Postdoctoral 
Fellowship Award Number PF~9-10007, issued by the Chandra X-ray 
Center, which is operated by the Smithsonian Astrophysical Observatory 
for and on behalf of NASA under contract NAS~8-39073. This work was
partially supported by NASA grant NAG~5-7980.

\newpage

\newpage

\newpage

\newpage

\newpage

\newpage

\newpage

\begin{table}
  \begin{center}
    \caption{Fit parameters to compute \gamav\ as a function of the plasma
             parameters using Eq. \ref{gamav_fit}. $\gamma_2 = 10^3$ has
             been used.}\vspace{1em}
    \renewcommand{\arraystretch}{1.2}
    \begin{tabular}[h]{cccccc}
      \hline
      $a_{\rm mxwl}$ & $p$ & $K_1$ & $\kappa_1$ & $K_2$ & $\kappa_2$ \\
      \hline 
0.28 & 2.20 & 2.326 & 0.60 & 2.329 & 3.098 \\
0.28 & 2.48 & 1.530 & 0.45 & 1.332 & 3.563 \\
0.28 & 2.73 & 0.661 & 0.30 & 0.876 & 1.013 \\
0.28 & 3.00 & 0.575 & 0.35 & 0.662 & 0.881 \\
0.28 & 3.24 & 0.405 & 0.40 & 0.576 & 0.766 \\
0.28 & 3.50 & 0.465 & 0.10 & 0.249 & 0.381 \\
0.44 & 2.20 & 1.759 & 0.45 & 1.761 & 0.766 \\
0.44 & 2.48 & 1.330 & 0.40 & 1.007 & 0.766 \\
0.44 & 2.73 & 2.023 & 0.75 & 0.761 & 2.037 \\
0.44 & 3.00 & 1.759 & 0.75 & 0.576 & 3.563 \\
0.44 & 3.24 & 1.530 & 0.70 & 0.435 & 3.563 \\
0.44 & 3.50 & 1.330 & 0.75 & 0.379 & 2.694 \\
0.60 & 2.20 & 2.675 & 0.75 & 1.332 & 0.664 \\
0.60 & 2.48 & 2.675 & 0.80 & 0.761 & 0.766 \\
0.60 & 2.73 & 2.675 & 0.80 & 0.501 & 1.540 \\
0.60 & 3.00 & 2.326 & 0.80 & 0.379 & 2.037 \\
0.60 & 3.24 & 2.023 & 0.85 & 0.329 & 1.540 \\
0.60 & 3.50 & 1.330 & 0.70 & 0.249 & 0.579 \\
0.76 & 2.20 & 2.326 & 0.70 & 0.761 & 0.503 \\
0.76 & 2.48 & 2.023 & 0.75 & 0.435 & 0.438 \\
0.76 & 2.73 & 2.326 & 0.90 & 0.329 & 0.579 \\
0.76 & 3.00 & 2.326 & 0.95 & 0.249 & 0.666 \\
0.76 & 3.24 & 2.023 & 0.90 & 0.188 & 0.579 \\
0.76 & 3.50 & 2.326 & 0.95 & 0.164 & 2.694 \\
0.92 & 2.20 & 2.675 & 0.95 & 0.249 & 0.579 \\
0.92 & 2.48 & 2.675 & 1.00 & 0.164 & 1.771 \\
0.92 & 2.73 & 2.326 & 1.00 & 0.108 & 0.766 \\
0.92 & 3.00 & 2.326 & 1.00 & 0.081 & 3.563 \\
0.92 & 3.24 & 1.759 & 0.95 & 0.062 & 0.331 \\
0.92 & 3.50 & 2.023 & 1.00 & 0.054 & 0.766 \\     
      \hline \\
      \end{tabular}
    \label{gm1_table}
  \end{center}
\end{table}

\end{document}